\documentclass[11pt]{revtex4}
\usepackage{mathtools}
\usepackage{hyperref}
\usepackage{pdflscape}
\usepackage{graphics}
\usepackage{subfigure}
\usepackage{epsfig}
\usepackage{amsmath,amsfonts}
\usepackage{amssymb}
\usepackage{color}
\usepackage{ulem}
\begin{document}
\title{Constraining Palatini-Horndeski theory with gravitational waves after GW170817}

\author{Yu-Qi Dong$^{a,b}$}
\email{dongyq21@lzu.edu.cn}

\author{Yu-Qiang Liu$^{a,b}$}
\email{liuyq18@lzu.edu.cn}

\author{Yu-Xiao Liu$^{a,b}$}
\email{liuyx@lzu.edu.cn, corresponding author}

\affiliation{$^{a}$Lanzhou Center for Theoretical Physics,
	Key Laboratory of Theoretical Physics of Gansu Province,
	School of Physical Science and Technology,
	Lanzhou University, Lanzhou 730000, China \\
	$^{b}$Institute of Theoretical Physics \& Research Center of Gravitation,
	Lanzhou University, Lanzhou 730000, China}

\begin{abstract}
\textbf{Abstract:} In this paper, we investigate the possible parameter space of Palatini-Horndeski theory with gravitational waves in a spatially flat Universe.
We find that if the theory 
satisfies the following condition: 
in any spatially flat cosmological background, the tensor gravitational wave speed is the speed of light $c$, then only $S = \int d^4x \sqrt{-g}  \big[K(\phi,X)-G_{3}(\phi,X){\tilde{\Box}}\phi+G_{4}(\phi)\tilde{R}\big]$ is left as the possible action in Palatini-Horndeski theory.
\end{abstract}
	
\maketitle

\section{Introduction}
\label{sec: intro}
The successful detection of gravitational waves made up the last puzzle missing in the experimental verification of general relativity \cite{Abbott1,Abbott2,Abbott3,Abbott4,Abbott5}. Therefore, general relativity has become the most successful theory of gravity so far.

However, there are still many theoretical problems that can not be explained by general relativity, such as how to explain the hierarchy between the Planck scale and the electroweak scale \cite{NimaArkani-Hamed,LRandall1,LRandall2} and how to quantize gravity \cite{A.Shomer}. In addition, the phenomena observed in experiments, such as the accelerated expansion of the Universe \cite{A.G.Riess} and the flat rotation curves of galaxies \cite{V.C.Rubin}, can not be explained by general relativity. For these reasons, many modified theories of gravity were considered \cite{NimaArkani-Hamed,LRandall1,LRandall2,Brans.Dicke,Horndeski,H.A.Buchdahl,C.Skordis,Timothy Clifton} in the hope of answering the problems that general relativity could not answer.

A well-defined modified theory of gravity should be stable. Ostrogradsky's research pointed out that when a Lagrangian contains higher-order (second order or higher) time derivatives of dynamical variables, its corresponding Hamiltonian is usually bilateral unbounded \cite{M.Ostrogradsky,R.P.Woodard}. It is generally believed that this unbounded Hamiltonian will lead to an instability of the theory called the Ostrogradsky instability \cite{R.P.Woodard,H.Motohashi,A.Ganz}. Therefore, a modified gravity theory with the Ostrogradsky instability is generally considered to be pathological and should be avoided.

Adding additional scalar field is one way to modify gravity. Theories obtained in this way are called scalar-tensor theories. In Ref. \cite{Z.Chen}, a tentative indication for scalar transverse gravitational waves was reported. If this is further confirmed in the future, it will strongly suggest that the gravity theory describing our world should have a scalar degree of freedom. In order to avoid the Ostrogradsky instability, we expect to give priority to those theories that can derive second-order field equations. In the metric formalism, the most general scalar-tensor theory that can derive second-order field equations is Horndeski theory \cite{Horndeski}.

However, Refs. \cite{P.Creminelli,C.D.Kreisch,Y.Gong} pointed out that the observation of the speed of tensor gravitational waves in the Universe by the gravitational wave event GW170817 together with the gamma ray burst GRB170817A would severely constrain the possible parameter space of metric Horndeski theory. Specifically, GW170817 and GRB170817A require the tensor gravitational wave speed $c_g$ to meet \cite{B.P.Abbott000,B.P.Abbott111}
\begin{eqnarray} 	
	\label{cg}
	-3\times10^{-15} \leq \frac{c_g}{c}-1 \leq 7\times10^{-16}.
\end{eqnarray}
This shows that in a very high precision, we can say that the tensor gravitational wave speed in the Universe is equal to the basic constant $c$ (speed of light). Considering that the cosmic background is also evolving during gravitational wave propagation, the most economic and natural assumption made by this observation result for the theory seems to be that: in any cosmological background, tensor gravitational waves always propagate at the speed of light. However, the possible subclasses of metric Horndeski theory satisfying this assumption only remain~\cite{P.Creminelli,R.Kase}
\begin{eqnarray} 	
 S=\int d^4x \sqrt{-g} \big[ K(\phi,X)-G_{3}(\phi,X){\Box}\phi+G_{4}(\phi)R \big].
\end{eqnarray}
This constraint limits the
application of scalar-tensor theories. Therefore, We expect to find scalar-tensor theories beyond the metric Horndeski framework. There are also many studies using GW170817 to constrain modify gravity theories~\cite{Y.Gong1,A.Gumrukcuoglu,Y.Gong2,Y.Cai,J.Oost,Y.Gong3,L.Shao}.

Further analysis shows that not all higher derivative theories have the Ostrogradsky instability. The higher derivative theory without the Ostrogradsky instability is required to satisfy the degeneracy condition \cite{H. Motohashi2,A.Ganz,D.Langlois}. In the metric formalism, the scalar-tensor theory with higher derivative but without the Ostrogradsky instability is called degenerate higher-order scalar-tensor (DHOST) theory \cite{D. Langlois,J.BenAchour,D.Langlois,J.Gleyzes,M.Crisostomi1,T. Kobayashi}. In addition to DHOST theory, considering the teleparallel framework is another way to go beyond the metric Horndeski framework. In teleparallel Horndeski theory established by Bahamonde \textit{et al}., metric Horndeski theory is included in the teleparallel framework as one of many subclasses \cite{S.Bahamonde1,S.Bahamonde2}.

Considering the scalar-tensor theory in the Palatini formalism may be another way to go beyond metric Horndeski framework. There have been some works on scalar-tensor gravity in the Palatini formalism \cite{U.Lindstrom,F.Bauer,M.Li,T.Markkanen,L.Jarv,K.Aoki1,A.Kozak,K.Shimada,R.Jinno,K.Aoki2,Helpin,Helpin2,M.Kubota,Y.Dong}. Cosmology in Palatini-Horndeski theory is different from that in metric Horndeski theory and their stability properties are different \cite{Helpin}. Different from metric Horndeski theory, under some parameter spaces, the connection of Palatini-Horndeski theory will introduce some new degrees of freedom \cite{Helpin}. In addition, the polarization modes of gravitational waves in Palatini-Horndeski theory are different from that in metric Horndeski theory \cite{Y.Dong}. Thus, it seems that Palatini-Horndeski theory may be different from metric Horndeski. However, it is necessary to further investigate the possible parameter space of Palatini-Horndeski theory.

In this paper, we will find possible subclasses of Palatini-Horndeski theory that 
 satisfied the following 
 condition: 
  the speed of tensor gravitational waves is the speed of light in any spatially flat cosmological background. 
 In Sec. \ref{sec: 2}, we will review Palatini-Horndeski theory. In Sec. \ref{sec: 3}, we will discuss the Ostrogradsky instability in Palatini-Horndeski theory for the evolution of a spatially flat Universe. 
 In Sec. \ref{sec: 4}, we will obtain the speed of tensor gravitational waves in the spatially flat cosmological background and 
 	constrain the parameter space. 
 	The conclusion will be given in Sec. \ref{sec: 5}.

We will use the natural system of units in this paper. Greek alphabet indices $(\mu,\nu,\lambda,\rho)$ and Latin alphabet indices $(i,j,k,l)$ range over spacetime indices $(0,1,2,3)$ and space indices $(1,2,3)$, respectively.

\section{Palatini-Horndeski theory}
\label{sec: 2}
In the Palatini formalism, the connection is independent of the metric. Therefore, it is necessary to take the variations of the action with respect to the metric and the connection independently. The Riemann tensor $\tilde{R}^{\mu}_{\ \nu\rho\sigma}$ and Ricci tensor $\tilde{R}_{\mu\nu}$ in the Palatini formalism are defined as
\begin{eqnarray}
	\label{Riemann}
	\tilde{R}^{\rho}_{~\mu\lambda\nu}
		&=&\partial_{\lambda}\Gamma^{\rho}_{\mu\nu}
		-\partial_{\nu}\Gamma^{\rho}_{\mu\lambda}
		+\Gamma^{\rho}_{\sigma\lambda}\Gamma^{\sigma}_{\mu\nu}
		-\Gamma^{\rho}_{\sigma\nu}\Gamma^{\sigma}_{\mu\lambda},
	\\	
	\label{Ricci}
	\tilde{R}_{\mu\nu}~~
	    &=&\partial_{\lambda}\Gamma^{\lambda}_{\mu\nu}
	    -\partial_{\nu}\Gamma^{\lambda}_{\mu\lambda}
	    +\Gamma^{\lambda}_{\sigma\lambda}\Gamma^{\sigma}_{\mu\nu}
	    -\Gamma^{\lambda}_{\sigma\nu}\Gamma^{\sigma}_{\mu\lambda}.
\end{eqnarray}
Furthermore, we assume that the connection is nontorsion: $\Gamma^{\lambda}_{\mu\nu}=\Gamma^{\lambda}_{\nu\mu}$.

The action of Palatini-Horndeski theory is defined as follows:
\begin{eqnarray}
	\label{action}
	S\; = \; \int d^4x \sqrt{-g}
	\;\Bigl(\mathcal{L}_{2}+\mathcal{L}_{3}+\mathcal{L}_{4}+\mathcal{L}_{5}\Bigl)
	\label{actioneq},
\end{eqnarray}
where
\begin{eqnarray}
	\label{L2}
	\mathcal{L}_{2}&=&K(\phi,X),
	\\
	\label{L3}
	\mathcal{L}_{3}&=&-G_{3}(\phi,X)\tilde{\Box}\phi,
	\\
	\label{L4}
	\mathcal{L}_{4}&=&G_{4}(\phi,X)\tilde{R}
	                +G_{4,X}(\mathnormal{\phi},X)
	                        \left[\left({\tilde{\Box}\phi}\right)^{2}
	                              -\left(\tilde{\nabla}_{\mu}\tilde{\nabla}_{\nu}\phi\right)
	                              \left(\tilde{\nabla}^{\mu}\tilde{\nabla}^{\nu}\phi\right)\right],
	\\
	\label{L5}
	\nonumber
	\mathcal{L}_{5}&=&G_{5}(\phi,X)\left(\tilde{R}_{\mu\nu}
	                -\frac{1}{2}\mathnormal{g}_{\mu\nu}\tilde{R}\right)
	                \tilde{\nabla}^{\mu}\tilde{\nabla}^{\nu}\phi
	\\ \nonumber
	&-&\frac{1}{6}G_{5,X}(\phi,X)
	\left[
	     \left(\tilde{\Box}\phi\right)^{3}-3\tilde{\Box}\phi
	     \left(\tilde{\nabla}_{\mu}\tilde{\nabla}_{\nu}\phi\right)
	     \left(\tilde{\nabla}^{\mu}\tilde{\nabla}^{\nu}\phi\right)\right.
	\\
	&+&2\left.\left(\tilde{\nabla}^{\lambda}\tilde{\nabla}_{\rho}\phi\right)
	    \left(\tilde{\nabla}^{\rho}\tilde{\nabla}_{\sigma}\phi\right)
	    \left(\tilde{\nabla}^{\sigma}\tilde{\nabla}_{\lambda}\phi\right)
	\right].
\end{eqnarray}
Here, $\tilde{\Box}=\tilde{\nabla}^{\mu}\tilde{\nabla}_{\mu}$, $X=-\frac{1}{2}\partial_{\mu}\phi\partial^{\mu}\phi$, and $ K, G_{3}, G_{4}$ and $G_{5}$ are real analytic functions of the variables $\phi$ and $X$. To distinguish the quantities in the metric formalism, we add tilde to represent the corresponding quantities defined in the Palatini formalism. A comma in subscript means partial derivative, e.g., $G_{4,X}\equiv \partial G_4 /\partial X$.

In the Palatini formalism, the compatibility condition $\tilde{\nabla}_{\lambda}g_{\mu\nu}=0$ is generally no longer valid. Therefore, the definition of $\tilde{\Box}\phi, \tilde{\nabla}^{\mu}\tilde{\nabla}^{\nu}\phi$ and $ \tilde{\nabla}^{\mu}\tilde{\nabla}_{\nu}\phi$ in the action (\ref{action}) in the Palatini formalism is not unique \cite{Helpin,M.Kubota}. In this paper, we take the definition in Ref. \cite{Y.Dong}:
\begin{eqnarray}	
	\nonumber	\label{nabla define}
	\tilde{\Box}\phi
	&=& g^{\mu\nu}\tilde{\nabla}_{\mu}\tilde{\nabla}_{\nu}\phi,
	\\
	\tilde{\nabla}^{\mu}\tilde{\nabla}^{\nu}\phi
	&=& g^{\mu\rho}\tilde{\nabla}_{\rho}(g^{\nu\sigma}\tilde{\nabla}_{\sigma}\phi),
	\\
	 \tilde{\nabla}^{\mu}\tilde{\nabla}_{\nu}\phi
	 &=& g^{\mu\rho}\tilde{\nabla}_{\rho}\tilde{\nabla}_{\nu}\phi. \nonumber
\end{eqnarray}

\section{The Ostrogradsky instability}
\label{sec: 3}
Since the action (\ref{action}) of Palatini-Horndeski theory contains the second order time derivatives of the scalar field $\phi$, one may think that Palatini-Horndeski theory has the Ostrogradsky instability. However, in this section, the analysis of the degeneracy condition in the case of the evolution of a spatially flat Universe shows that it can not be taken for granted that  Palatini-Horndeski theory must have the Ostrogradsky instability.

If a theory is Ostrogradsky stable, then it must be Ostrogradsky stable in the special case of the evolution of a spatially flat Universe. 
For a spatially flat Universe, the metric $g_{\mu\nu}$ is the spatially flat Friedmann-Robertson-Walker (FRW) metric, and the connection $\Gamma^{\lambda}_{\mu\nu}$ and scalar field $\phi$ are only functions of time:
\begin{eqnarray}	
	\label{background}
	ds^2=-N(t)^2 dt^2+a(t)^2\delta_{ij}dx^idx^j, \quad
	\Gamma^{\lambda}_{\mu\nu}=\Gamma^{\lambda}_{\mu\nu}(t), \quad
	\phi=\phi(t).
\end{eqnarray}
We consider that the connection has spatial isotropy, that is, under the spatial rotation transformation, the components of the connection $\Gamma^{\lambda}_{\mu\nu}$ are invariant. This condition further limits the value of the connection. Specifically, under the spatial rotation transformation, the transformation law of the components of the connection is the same as that of the third-order tensor, which requires that the connection $\Gamma^{\lambda}_{\mu\nu}$ satisfies \cite{A.Minkevich,D.Iosifidis}
\begin{eqnarray}	
	\label{Gamma background}
	&\Gamma^{0}_{0i}=\Gamma^{i}_{00}=0, \quad
	\Gamma^{0}_{ij}=\Gamma^{0}_{11} \delta_{ij}, \quad
	\Gamma^{i}_{0j}=\Gamma^{1}_{01} \delta^{i}_{j}, \nonumber \\
	&\Gamma^{i}_{jk}
	    =\Gamma^{1}_{23} \delta^{il} \varepsilon_{ljk}
    	=\Gamma^{1}_{23} \delta^{il}  \varepsilon_{l(jk)}
    	=0.
\end{eqnarray}
Here, $\delta_{ij}$ is the Kronecker delta, and $\varepsilon_{ijk}$ is the Levi-Civita tensor. It can be seen that for the components of the connection, only $\Gamma^{0}_{00}, \Gamma^{0}_{11}=\Gamma^{0}_{22}=\Gamma^{0}_{33}$ and $\Gamma^{1}_{01}=\Gamma^{2}_{02}=\Gamma^{3}_{03}$ may not be zero.

One may want to set $N(t)=1$ at the level of action, and then obtain the evolution equations by varying the variables $\left(a,\phi,\Gamma^{0}_{00},\Gamma^{0}_{11},\Gamma^{1}_{01}\right)$. However, this will miss one equation \cite{H.MotohashiT.SuyamaK.Takahashi}. Thus, in order to obtain complete evolution equations, $N$ should be kept in the action.

By substituting Eqs. (\ref{background}) and (\ref{Gamma background}) into the action (\ref{action}), we obtain the action that describes the evolution of a spatially flat Universe:
\begin{eqnarray}
	\label{Universe action}
	S\; = \; \int dt\int d^3x~ L\left(\ddot{\phi},\dot{\phi},\phi,\dot{N},N,a,\Gamma^{0}_{00},\dot{\Gamma}^{0}_{11},{\Gamma}^{0}_{11},\dot{\Gamma}^{1}_{01},{\Gamma}^{1}_{01}\right),
\end{eqnarray}
where
\begin{eqnarray}
	\label{L}
	\frac{L}{a^3 N}&=&
	  	K
	   +\frac{3G_3\Gamma^{0}_{11}\dot{\phi}}{a^2}
	   -\frac{G_3}{N^2}\left(\Gamma^{0}_{00}\dot{\phi}-\ddot{\phi}\right)
   	   \nonumber \\
   	  &+&
   	  	\frac{3G_4}{a^2}\left(
   	  	\Gamma^{0}_{00}\Gamma^{0}_{11}+\Gamma^{0}_{11}\Gamma^{1}_{01}+\dot{\Gamma}^{0}_{11}
   	  	\right)
   	  -\frac{3G_4}{N^2}\left(
   	  	\Gamma^{0}_{00}\Gamma^{1}_{01}-{\Gamma^{1}_{01}}^2-\dot{\Gamma}^{1}_{01}
   	  	\right)
      \nonumber \\
      &+&
      	\frac{3G_5\Gamma^{1}_{01}\dot{\phi}}{2a^2N^4}
                  \left[
                      N^2\left(
                        \Gamma^{0}_{00}\Gamma^{0}_{11}+\Gamma^{0}_{11}\Gamma^{1}_{01}+\dot{\Gamma}^{0}_{11}
                         \right)
                      +3a^2\left(
                             -\Gamma^{0}_{00}\Gamma^{1}_{01}+{\Gamma^{1}_{01}}^2+\dot{\Gamma}^{1}_{01}
                           \right)
                  \right]
	\nonumber \\
	&+&
		\frac{3G_5}{2a^2N^5}
	       \left[
	          N^2\left(
	               \Gamma^{0}_{00}\Gamma^{0}_{11}+\Gamma^{0}_{11}\Gamma^{1}_{01}+\dot{\Gamma}^{0}_{11}
	             \right)
	         +a^2\left(
	               \Gamma^{0}_{00}\Gamma^{1}_{01}-{\Gamma^{1}_{01}}^2-\dot{\Gamma}^{1}_{01}
	             \right)
	       \right]
	\nonumber \\
	&\times&
	\left[
	  -2\dot{N}\dot{\phi}+N\left(\Gamma^{0}_{00}\dot{\phi}+\ddot{\phi}\right)
	\right]
	+\frac{G_{4,X}\dot{\phi}}{a^4N^5}
	\left[
	9N^5{\Gamma^{0}_{11}}^2\dot{\phi}
	\right.
	\nonumber \\
	&-&
	           \left.
	3a^2N^3\Gamma^{0}_{11}
	                      \left(
	                        2\Gamma^{0}_{00}\dot{\phi}+\Gamma^{1}_{01}\dot{\phi}-2\ddot{\phi}
	                      \right)
	                +2a^4\left(
	                       N\Gamma^{0}_{00}-\dot{N}
 	                     \right)
 	                     \left(
 	                     \Gamma^{0}_{00}\dot{\phi}-\ddot{\phi}
 	                     \right)  	
	             \right]
	\nonumber \\
	&-&
		\frac{G_{5,X}}{2a^6N^7}\dot{\phi}
	     \left[
	       -11N^7{\Gamma^{0}_{11}}^3\dot{\phi}^2-3a^4N^2\Gamma^{0}_{11}
	             \left(
	                N\left(2\Gamma^{0}_{00}+\Gamma^{1}_{01}\right)-2\dot{N}
	             \right)
	          \dot{\phi}\left(\Gamma^{0}_{00}\dot{\phi}-\ddot{\phi}\right)
	     \right.
	\nonumber \\
	&+&
	\left.
	   9a^2N^5{\Gamma^{0}_{11}}^2\dot{\phi}
	     \left(
	       \Gamma^{0}_{00}\dot{\phi}+\Gamma^{1}_{01}\dot{\phi}-\ddot{\phi}
	     \right)
	   +2a^6\left(N\Gamma^{0}_{00}-\dot{N}\right)\left(-\Gamma^{0}_{00}\dot{\phi}+\ddot{\phi}\right)^2
	   \right].
\end{eqnarray}
Here and below, the dot on the letter represents the derivative of the corresponding quantity with respect to time, $K,G_3,G_4,G_5,G_{4,X}$, and $G_{5,X}$ are functions of $(\phi,\frac{\dot{\phi}^2}{2N^2})$. Because $L$ is only a function of $t$ in current case, we can now omit $\int d^3x$ and consider $L$ itself as a Lagrangian.

By using a Lagrangian multiplier $\pi$ to impose the constraint $\dot{\phi}=s$, the Lagrangian $L$ in (\ref{L}) can be equivalent to the following Lagrangian $\tilde{L}$, which only includes dynamic variables and the first order time derivative of dynamic variables:
\begin{eqnarray}
	\label{eq L}
	&\tilde{L}\left(\dot{s},s,\dot{\phi},\phi,\dot{N},N,a,\pi,\Gamma^{0}_{00},\dot{\Gamma}^{0}_{11},{\Gamma}^{0}_{11},\dot{\Gamma}^{1}_{01},{\Gamma}^{1}_{01}\right) \nonumber\\
	&=L\left(\dot{s},s,\phi,\dot{N},N,a,\Gamma^{0}_{00},\dot{\Gamma}^{0}_{11},{\Gamma}^{0}_{11},\dot{\Gamma}^{1}_{01},{\Gamma}^{1}_{01}\right)+\pi \left(\dot{\phi}-s\right).
\end{eqnarray}
By varying the Lagrangian (\ref{eq L}), two kinds of equations can be obtained. The first kind of equations  include
\begin{eqnarray}
	\label{dl/dpi}
   \frac{\partial{\tilde{L}}}{\partial{\pi}}
   &=&\dot{\phi}-s
   =0, 	\\
	\label{dl/da}
	\frac{\partial{\tilde{L}}}{\partial{a}}	&=&\frac{\partial{L}}{\partial{a}}
           \left(\dot{s},s,\phi,\dot{N},N,a,\Gamma^{0}_{00},\dot{\Gamma}^{0}_{11},
                {\Gamma}^{0}_{11},\dot{\Gamma}^{1}_{01},{\Gamma}^{1}_{01}\right)
	=0,	\\
	\label{dl/dgamma000}
	\frac{\partial{\tilde{L}}}{\partial{\Gamma^{0}_{00}}}	
   &=&\frac{\partial{L}}{\partial{\Gamma^{0}_{00}}}
       \left(\dot{s},s,\phi,\dot{N},N,a,\Gamma^{0}_{00},\dot{\Gamma}^{0}_{11},
           {\Gamma}^{0}_{11},\dot{\Gamma}^{1}_{01},{\Gamma}^{1}_{01}\right)
	=0.
\end{eqnarray}
They are constraints between variables $\left(\dot{s},s,\phi,\dot{N},N,a,\Gamma^{0}_{00},\dot{\Gamma}^{0}_{11},{\Gamma}^{0}_{11},\dot{\Gamma}^{1}_{01},{\Gamma}^{1}_{01}\right)$. The second kind of equations are
\begin{eqnarray}
	\frac{d}{dt}\frac{\partial{\tilde{L}}}{\partial{\dot{N}}}-\frac{\partial{\tilde{L}}}{\partial{N}}
	=0,\quad
	\frac{d}{dt}\frac{\partial{\tilde{L}}}{\partial{\dot{s}}}-\frac{\partial{\tilde{L}}}{\partial{s}}
	=0,\quad
	\frac{d}{dt}\frac{\partial{\tilde{L}}}{\partial{\dot{\phi}}}
    -\frac{\partial{\tilde{L}}}{\partial{\phi}}
	=0,\label{E-L eq1}	\\	
	\frac{d}{dt}\frac{\partial{\tilde{L}}}{\partial{\dot{\Gamma}^{0}_{11}}}
     -\frac{\partial{\tilde{L}}}{\partial{{\Gamma}^{0}_{11}}}
	=0,\quad
	\frac{d}{dt}\frac{\partial{\tilde{L}}}{\partial{\dot{\Gamma}^{1}_{01}}}
     -\frac{\partial{\tilde{L}}}{\partial{{\Gamma}^{1}_{01}}}
	=0.\qquad\quad \label{E-L eq2}
\end{eqnarray}
They are Euler-Lagrange equations. In addition, canonical momentums are defined as
\begin{eqnarray}
	\label{Pphi}
	P_{\phi}&=&\frac{\partial{\tilde{L}}}{\partial{\dot{\phi}}}
	= \pi,
	\\
	\label{PN}
	P_{N}&=&\frac{\partial{\tilde{L}}}{\partial{\dot{N}}}
	=\frac{\partial{L}}{\partial{\dot{N}}}\left(\dot{s},s,\phi,\dot{N},N,a,\Gamma^{0}_{00},\dot{\Gamma}^{0}_{11},{\Gamma}^{0}_{11},\dot{\Gamma}^{1}_{01},{\Gamma}^{1}_{01}\right),
	\\
	\label{Ps}
	P_{s}&=&\frac{\partial{\tilde{L}}}{\partial{\dot{s}}}
	=\frac{\partial{L}}{\partial{\dot{s}}}\left(\dot{s},s,\phi,\dot{N},N,a,\Gamma^{0}_{00},\dot{\Gamma}^{0}_{11},{\Gamma}^{0}_{11},\dot{\Gamma}^{1}_{01},{\Gamma}^{1}_{01}\right),
	\\
	\label{PGamma011}
	P_{\Gamma^{0}_{11}}&=&\frac{\partial{\tilde{L}}}{\partial{\dot{\Gamma}^{0}_{11}}}
	=\frac{\partial{L}}{\partial{\dot{\Gamma}^{0}_{11}}}\left(\dot{s},s,\phi,\dot{N},N,a,\Gamma^{0}_{00},\dot{\Gamma}^{0}_{11},{\Gamma}^{0}_{11},\dot{\Gamma}^{1}_{01},{\Gamma}^{1}_{01}\right),
	\\
	\label{PGamma101}
	P_{\Gamma^{1}_{01}}&=&\frac{\partial{\tilde{L}}}{\partial{\dot{\Gamma}^{1}_{01}}}
	=\frac{\partial{L}}{\partial{\dot{\Gamma}^{1}_{01}}}\left(\dot{s},s,\phi,\dot{N},N,a,\Gamma^{0}_{00},\dot{\Gamma}^{0}_{11},{\Gamma}^{0}_{11},\dot{\Gamma}^{1}_{01},{\Gamma}^{1}_{01}\right).
\end{eqnarray}

In order to analyze the Ostrogradsky stability, it is necessary to introduce the Hamiltonian formalism of the theory. For this, take the total differential of $\tilde{L}$
\begin{eqnarray}
	\label{dL}	
	d\tilde{L}
	&=&\frac{\partial{\tilde{L}}}{\partial{\dot{s}}} d{\dot{s}}
	+\frac{\partial{\tilde{L}}}{\partial{{s}}} d{{s}}
	+\frac{\partial{\tilde{L}}}{\partial{\dot{\phi}}} d{\dot{\phi}}
	+\frac{\partial{\tilde{L}}}{\partial{{\phi}}}d{{\phi}}
	+\frac{\partial{\tilde{L}}}{\partial{\dot{N}}} d{\dot{N}}
	+\frac{\partial{\tilde{L}}}{\partial{{N}}} d{{N}}
	+\frac{\partial{\tilde{L}}}{\partial{\dot{\Gamma}^{0}_{11}}} d{\dot{\Gamma}^{0}_{11}}
	 \nonumber
    \\
    &+&\frac{\partial{\tilde{L}}}{\partial{{{\Gamma}^{0}_{11}}}} d{{{\Gamma}^{0}_{11}}}
    +\frac{\partial{\tilde{L}}}{\partial{\dot{\Gamma}^{1}_{01}}}d\dot{\Gamma}^{1}_{01}
    +\frac{\partial{\tilde{L}}}{\partial{{\Gamma}^{1}_{01}}} d{\Gamma}^{1}_{01}
    +\frac{\partial{\tilde{L}}}{\partial{\pi}} d\pi+\frac{\partial{\tilde{L}}}{\partial{a}} da
    +\frac{\partial{\tilde{L}}}{\partial{{\Gamma}^{0}_{00}}}d{\Gamma}^{0}_{00}.
\end{eqnarray}
Using Eqs.~(\ref{dl/dpi})-(\ref{E-L eq2}) and the definitions of canonical momentums (\ref{Pphi})-(\ref{PGamma101}), the expression (\ref{dL}) will be equivalent to
\begin{eqnarray}
	\label{dH}	
   && d\left(P_s\dot{s}+P_\phi\dot{\phi}+P_N\dot{N}+P_{{\Gamma}^{0}_{11}}\dot{\Gamma}^{0}_{11}
          +P_{{\Gamma}^{1}_{01}}\dot{\Gamma}^{1}_{01}-\tilde{L}
    \right) \nonumber	\\	
  &=&\dot{s}dP_{s}
	-\dot{P_{s}}ds
	+\dot{\phi}dP_{\phi}
	-\dot{P_{\phi}}d\phi
	+\dot{N}dP_{N}
	-\dot{P}_{N}dN  \nonumber	\\	
   &+&\dot{\Gamma}^{0}_{11}dP_{{{\Gamma}^{0}_{11}}}
    -\dot{P}_{{\Gamma}^{0}_{11}}d{{\Gamma}^{0}_{11}}
	+\dot{\Gamma}^{1}_{01}dP_{{{\Gamma}^{1}_{01}}}
	-\dot{P}_{{\Gamma}^{1}_{01}}d{{\Gamma}^{1}_{01}}.
\end{eqnarray}
It allows that the Hamiltonian of the theory is defined as
\begin{eqnarray}
	\label{H}
	H=P_s\dot{s}+P_\phi\dot{\phi}+P_N\dot{N}+P_{{\Gamma}^{0}_{11}}\dot{\Gamma}^{0}_{11}+P_{{\Gamma}^{1}_{01}}\dot{\Gamma}^{1}_{01}-\tilde{L}.
\end{eqnarray}
If we can use the first kind of equations (\ref{dl/dpi})-(\ref{dl/dgamma000}) and the definitions of canonical momentums (\ref{Pphi})-(\ref{PGamma101}) to express variables $\left(\dot{s},\dot{\phi},\dot{N},\dot{\Gamma}^{0}_{11},\dot{\Gamma}^{1}_{01},{\Gamma}^{0}_{00},a,\pi\right)$ as functions of independent variables $\left(s,P_{s},\phi,P_{\phi},N,P_{N},{\Gamma}^{0}_{11},P_{{\Gamma}^{0}_{11}},{\Gamma}^{1}_{01},P_{{\Gamma}^{1}_{01}}\right)$, so as to express the Hamiltonian $H$ as a function of independent variables $\left(s,P_{s},\phi,P_{\phi},N,P_{N},{\Gamma}^{0}_{11},P_{{\Gamma}^{0}_{11}},{\Gamma}^{1}_{01},P_{{\Gamma}^{1}_{01}}\right)$, then using (\ref{dH}), we can obtain the Hamilton's equations equivalent to the Euler-Lagrange equations (\ref{E-L eq1})(\ref{E-L eq2}):
\begin{equation}
\label{Hamilton's equations}
        \begin{aligned}	
	\dot{s}&=\frac{\partial{H}}{\partial{P_{s}}},\quad~~~~~~
	\dot{P_s}=-\frac{\partial{H}}{\partial{{s}}}; \\
	\dot{\phi}&=\frac{\partial{H}}{\partial{P_{\phi}}},\quad ~~~~~\,
    \dot{P_\phi}=-\frac{\partial{H}}{\partial{{\phi}}};  \\
    \dot{N}&=\frac{\partial{H}}{\partial{P_{N}}},\quad~~~~~
    \dot{P}_{N}=-\frac{\partial{H}}{\partial{{N}}}; \\
	\dot{\Gamma}^{0}_{11}&=\frac{\partial{H}}{\partial{P_{\Gamma^{0}_{11}}}},\quad  ~
    \dot{P}_{\Gamma^{0}_{11}}=-\frac{\partial{H}}{\partial{{\Gamma}^{0}_{11}}};\\
	\dot{\Gamma}^{1}_{01}&=\frac{\partial{H}}{\partial{P_{\Gamma^{1}_{01}}}},\quad ~
    \dot{P}_{\Gamma^{1}_{01}}=-\frac{\partial{H}}{\partial{{\Gamma}^{1}_{01}}}.
\end{aligned}
\end{equation}

The implicit function theorem gives a sufficient condition for the following proposition
: variables $\left(\dot{s},\dot{\phi},\dot{N},\dot{\Gamma}^{0}_{11},\dot{\Gamma}^{1}_{01},{\Gamma}^{0}_{00},a,\pi \right)$ can be represented by variables $\left(s,P_{s},\phi,P_{\phi},N,P_{N},{\Gamma}^{0}_{11},P_{{\Gamma}^{0}_{11}},{\Gamma}^{1}_{01},P_{{\Gamma}^{1}_{01}}\right)$ locally. We mark all the set of variables $\left(\dot{s},\dot{\Gamma}^{0}_{11},\dot{\Gamma}^{1}_{01},\dot{N},{\Gamma}^{0}_{00},a,s,P_{s},\phi,N,P_{N},{\Gamma}^{0}_{11},P_{{\Gamma}^{0}_{11}},{\Gamma}^{1}_{01},P_{{\Gamma}^{1}_{01}}\right)$ as $\mathcal{X}$. This theorem points out that for a solution $x_0 \in \mathcal{X}$ of the variables satisfying Eqs.~(\ref{dl/da}), (\ref{dl/dgamma000}), and (\ref{PN})-(\ref{PGamma101}), if the value of $\mathcal{K}\left(\dot{s},s,\phi,\dot{N},N,a,\Gamma^{0}_{00},\dot{\Gamma}^{0}_{11},{\Gamma}^{0}_{11},\dot{\Gamma}^{1}_{01},{\Gamma}^{1}_{01}\right)$ is not zero at $x_0$, where
\begin{eqnarray}
	\label{K}
	\mathcal{K}
	=\Large {\begin{vmatrix}
			\frac{{\partial}^2 L}{{\partial{a}}^2} &
			\frac{{\partial}^2 L}{\partial{a}\partial{\dot{s}}} &
			\frac{{\partial}^2 L}{\partial{a}\partial{\dot{\Gamma}^{0}_{11}}}&
			\frac{{\partial}^2 L}{\partial{a}\partial{\dot{\Gamma}^{1}_{01}}}&
			\frac{{\partial}^2 L}{\partial{a}\partial{{\Gamma}^{0}_{00}}}&
			\frac{{\partial}^2 L}{\partial{a}\partial{\dot{N}}}
			\\
		    \frac{{\partial}^2 L}{\partial{\dot{s}}{\partial{a}}} &
		    \frac{{\partial}^2 L}{{\partial{\dot{s}}}^2} &
		    \frac{{\partial}^2 L}{\partial{\dot{s}}\partial{\dot{\Gamma}^{0}_{11}}}&
		    \frac{{\partial}^2 L}{\partial{\dot{s}}\partial{\dot{\Gamma}^{1}_{01}}}&
		    \frac{{\partial}^2 L}{\partial{\dot{s}}\partial{{\Gamma}^{0}_{00}}}&
		    \frac{{\partial}^2 L}{\partial{\dot{s}}\partial{\dot{N}}}
		    \\	
		    \frac{{\partial}^2 L}{\partial{\dot{\Gamma}^{0}_{11}}{\partial{a}}} &
			\frac{{\partial}^2 L}{\partial{\dot{\Gamma}^{0}_{11}}{\partial{\dot{s}}}} &
			\frac{{\partial}^2 L}{{\partial{\dot{\Gamma}^{0}_{11}}}^2}&
			\frac{{\partial}^2 L}{\partial{\dot{\Gamma}^{0}_{11}}\partial{\dot{\Gamma}^{1}_{01}}}&
			\frac{{\partial}^2 L}{\partial{\dot{\Gamma}^{0}_{11}}\partial{{\Gamma}^{0}_{00}}}&
			\frac{{\partial}^2 L}{\partial{\dot{\Gamma}^{0}_{11}}\partial{\dot{N}}}
			\\	
			\frac{{\partial}^2 L}{\partial{\dot{\Gamma}^{1}_{01}}{\partial{a}}} &
			\frac{{\partial}^2 L}{\partial{\dot{\Gamma}^{1}_{01}}{\partial{\dot{s}}}} &
			\frac{{\partial}^2 L}{\partial{\dot{\Gamma}^{1}_{01}}\partial{\dot{\Gamma}^{0}_{11}}}&
			\frac{{\partial}^2 L}{{\partial{\dot{\Gamma}^{1}_{01}}}^2}&
			\frac{{\partial}^2 L}{\partial{\dot{\Gamma}^{1}_{01}}\partial{{\Gamma}^{0}_{00}}}&
			\frac{{\partial}^2 L}{\partial{\dot{\Gamma}^{1}_{01}}\partial{\dot{N}}}
			 \\	
			\frac{{\partial}^2 L}{\partial{{\Gamma}^{0}_{00}}{\partial{a}}} &
			\frac{{\partial}^2 L}{\partial{{\Gamma}^{0}_{00}}{\partial{\dot{s}}}} &
			\frac{{\partial}^2 L}{\partial{{\Gamma}^{0}_{00}}\partial{\dot{\Gamma}^{0}_{11}}}&
			\frac{{\partial}^2 L}{\partial{{\Gamma}^{0}_{00}}\partial{\dot{\Gamma}^{1}_{01}}}&
			\frac{{\partial}^2 L}{{\partial{{\Gamma}^{0}_{00}}}^2}&
			\frac{{\partial}^2 L}{\partial{{\Gamma}^{0}_{00}}\partial{\dot{N}}}
			\\
		    \frac{{\partial}^2 L}{\partial{\dot{N}}\partial{a}} &
		    \frac{{\partial}^2 L}{\partial{\dot{N}}\partial{\dot{s}}}&
		    \frac{{\partial}^2 L}{\partial{\dot{N}}\partial{\dot{\Gamma}^{0}_{11}}}&
		    \frac{{\partial}^2 L}{\partial{\dot{N}}\partial{\dot{\Gamma}^{1}_{01}}}&
		    \frac{{\partial}^2 L}{\partial{\dot{N}}\partial{{\Gamma}^{0}_{00}}}&
		    \frac{{\partial}^2 L}{{\partial{\dot{N}}^2}}
	
\end{vmatrix}},
\end{eqnarray}
then there exists a neighbourhood $\mathcal{O}\subseteq\mathcal{X}$ of $x_0$, such that the following relationships can be solved for all points in $\mathcal{O}$ that satisfy Eqs.~(\ref{dl/da}), (\ref{dl/dgamma000}), and (\ref{Ps})-(\ref{PGamma101}):
\begin{eqnarray}
\dot{s}
    &=&\dot{s}\left(s,P_{s},\phi,N,P_{N},{\Gamma}^{0}_{11},P_{{\Gamma}^{0}_{11}},{\Gamma}^{1}_{01},P_{{\Gamma}^{1}_{01}}\right),
\nonumber
\\
a
    &=&a\left(s,P_{s},\phi,N,P_{N},{\Gamma}^{0}_{11},P_{{\Gamma}^{0}_{11}},{\Gamma}^{1}_{01},P_{{\Gamma}^{1}_{01}}\right),
\nonumber
\\
\dot{N}
&=&\dot{N}\left(s,P_{s},\phi,N,P_{N},{\Gamma}^{0}_{11},P_{{\Gamma}^{0}_{11}},{\Gamma}^{1}_{01},P_{{\Gamma}^{1}_{01}}\right),
\nonumber
\\
\Gamma^{0}_{00}
    &=&\Gamma^{0}_{00}\left(s,P_{s},\phi,N,P_{N},{\Gamma}^{0}_{11},P_{{\Gamma}^{0}_{11}},{\Gamma}^{1}_{01},P_{{\Gamma}^{1}_{01}}\right),
\label{sol}
\\
\dot{\Gamma}^{0}_{11}
    &=&\dot{\Gamma}^{0}_{11}\left(s,P_{s},\phi,N,P_{N},{\Gamma}^{0}_{11},P_{{\Gamma}^{0}_{11}},{\Gamma}^{1}_{01},P_{{\Gamma}^{1}_{01}}\right),
\nonumber
\\
\dot{\Gamma}^{1}_{01}
    &=&\dot{\Gamma}^{1}_{01}\left(s,P_{s},\phi,N,P_{N},{\Gamma}^{0}_{11},
    P_{{\Gamma}^{0}_{11}},{\Gamma}^{1}_{01},P_{{\Gamma}^{1}_{01}}\right). \nonumber
\end{eqnarray}
Thus, for variables satisfying Eqs.~(\ref{dl/dpi})-(\ref{dl/dgamma000}) and (\ref{PN})-(\ref{PGamma101}) in $\mathcal{O}$, using relationships (\ref{sol}), the Hamiltonian $H\left(s,P_{s},\phi,P_{\phi},N,P_{N},{\Gamma}^{0}_{11},P_{{\Gamma}^{0}_{11}},{\Gamma}^{1}_{01},P_{{\Gamma}^{1}_{01}}\right)$ can be locally expressed as
\begin{eqnarray}
	\label{H(p,q)}
	H
	&=&P_{\phi}s	+P_{s}\dot{s}\left(s,P_{s},\phi,N,P_{N},{\Gamma}^{0}_{11},P_{{\Gamma}^{0}_{11}},{\Gamma}^{1}_{01},P_{{\Gamma}^{1}_{01   }}\right)	\nonumber \\
  &+& P_{{\Gamma}^{0}_{11}}\dot{\Gamma}^{0}_{11}
     \left(s,P_{s},\phi,N,P_{N},{\Gamma}^{0}_{11},P_{{\Gamma}^{0}_{11}},{\Gamma   }^{1}_{01},P_{{\Gamma}^{1}_{01}}\right)
   \nonumber	\\	
   &+&P_{{\Gamma}^{1}_{01}}\dot{\Gamma}^{1}_{01}
    \left(s,P_{s},\phi,N,P_{N},{\Gamma}^{0}_{11},P_{{\Gamma}^{0}_{11}},{\Gamma   }^{1}_{01},P_{{\Gamma}^{1}_{01}}\right)  \nonumber	\\	
   	&-& L\left(s,P_{s},\phi,N,P_{N},{\Gamma}^{0}_{11},P_{{\Gamma}^{0}_{11}},
            {\Gamma}^{1}_{01},P_{{\Gamma}^{1}_{01}}\right).
\end{eqnarray}
Note that $P_{\phi}$ can take any real value and only appear in the  first term on the right hand side of Eq.~(\ref{H(p,q)}). Therefore, if we take the point in $\mathcal{O}$ that makes $s\neq0$, we can see that the Hamiltonian $H$ is bilateral unbounded, so the theory has the Ostrogradsky instability.

According to the above discussion, it can be seen that a necessary condition which we call degeneracy condition for Palatini-Horndeski theory to be Ostrogradsky stable is that the value of $\mathcal{K}$ at any variables $\left(\dot{s},s,\phi,\dot{N},N,a,\Gamma^{0}_{00},\dot{\Gamma}^{0}_{11},{\Gamma}^{0}_{11},\dot{\Gamma}^{1}_{01},{\Gamma}^{1}_{01}\right)$ satisfying Eqs. (\ref{dl/da}) and (\ref{dl/dgamma000}) is always $0$ \cite{A.Ganz}.

One might want to use this degenerate condition to rule out some unstable classes in Palatini-Horndeski theory. However, when substituting the Lagrangian (\ref{L}) into the definition of $\mathcal{K}$ in (\ref{K}), we are surprised to find
\begin{eqnarray}
	\label{K=0}
	\mathcal{K}=0.
\end{eqnarray}
This shows that all parameter spaces of Palatini-Horndeski satisfy the degeneracy condition.

Although this does not mean that all parameter spaces in the theory are Ostrogradsky stable, it shows that Palatini-Horndeski theory is not as easy to have the Ostrogradsky instability as expected. In fact, $\mathcal{K}=0$ means that there is at least one constraint on the phase space in the theory \cite{A.Ganz}. It is necessary to further analyze the constraint condition to clearly judge whether Palatini-Horndeski theory has the Ostrogradsky instability. Such an analysis seems very complex. However, we will easily see in Sec. \ref{sec: 5} that the parameter space of Palatini-Horndeski theory compatible with GW170817 is Ostrogradsky stable.

\section{The speed of tensor gravitational waves}
\label{sec: 4}


In this section, we will calculate the speed of tensor gravitational waves propagating in a spatially flat cosmological background 
and find possible subclasses of Palatini-Horndeski theory that 
satisfy the following 
condition: 
the speed of tensor gravitational waves is the speed of light in any spatially flat cosmological background.

In addition to the gravitational field, the ideal fluid material field is also distributed in the spatially flat Universe. Therefore, in addition to the gravitational field action (\ref{action}), we should also add an action $S_m$ that describes the ideal fluid into the the total action
\begin{eqnarray}
	\label{action tot}
	S_{tot}=S+S_m.
\end{eqnarray}
Here, $S$ is defined by (\ref{action}). In the Palatini formalism, $S_m$ is only a function of the metric and the material field, and it is independent of the connection. Varying the action $S_m$ with respect to $g_{\mu\nu}$, we obtain
\begin{eqnarray}
	\label{delta Sm}
	\delta S_m=-\frac{1}{2}\int d^4x \sqrt{-g} T^{\mu\nu} \delta{g_{\mu\nu}}.
\end{eqnarray}
Here, $T^{\mu\nu}$ is the energy-momentum tensor of the ideal fluid:
\begin{eqnarray}
	\label{T_munu}
	T^{\mu\nu}=(P+\epsilon)u^{\mu}u^{\nu}+Pg^{\mu\nu},
\end{eqnarray}
where $\epsilon$ is the matter density, $P$ is the matter pressure. The four-velocity $u^{\mu}$ satisfies $u^{0}=\frac{1}{N}, u^{i}=0$.

By substituting Eqs. (\ref{background}) and (\ref{Gamma background}) into the action (\ref{action tot}), and varying the action (\ref{action tot}) with respect to $N,a,\phi,\Gamma^{0}_{00},\Gamma^{0}_{11}$ and  $\Gamma^{1}_{01}$, we obtain the background equations:
\begin{eqnarray}
	\label{BG EQ with P}
  & \frac{d}{dt}\frac{\partial{{L}}}{\partial{\dot{N}}}-\frac{\partial{{L}}}{\partial{N}}+a^3\epsilon=0,\quad
  \frac{\partial{L}}{\partial{a}}-3a^2 P=0,\quad
   \frac{d^2}{{dt}^2}\frac{\partial{{L}}}{\partial{\ddot{\phi}}}
          -\frac{d}{dt}\frac{\partial{{L}}}{\partial{\dot{\phi}}}
          +\frac{\partial{{L}}}{\partial{{\phi}}}=0,    \nonumber
   \\
  & \frac{\partial{L}}{\partial{\Gamma^{0}_{00}}}=0, \quad
   \frac{d}{dt}\frac{\partial{{L}}}{\partial{\dot{\Gamma}^{0}_{11}}}
      -\frac{\partial{{L}}}{\partial{{\Gamma}^{0}_{11}}}=0, \quad
   \frac{d}{dt}\frac{\partial{{L}}}{\partial{\dot{\Gamma}^{1}_{01}}}
      -\frac{\partial{{L}}}{\partial{{\Gamma}^{1}_{01}}}=0.
\end{eqnarray}
Here, $L$ is defined by (\ref{L}). Because the specific expressions of the background equations (\ref{BG EQ with P}) are very lengthy and easy to obtain, they will not be listed in this paper. In the following, we take $N(t)=1$.

In order to study the tensor gravitational waves, we need to obtain the linear perturbation equations of the tensor perturbations on the spatially flat cosmological background.

Since the metric and connection are independent in the Palatini formalism, they should be perturbed independently:
\begin{eqnarray}
	\label{perturb}
	g_{\mu\nu} \rightarrow g_{\mu\nu}+h_{\mu\nu}, \quad
	\Gamma^{\lambda}_{\mu\nu} \rightarrow \Gamma^{\lambda}_{\mu\nu}+\Sigma^{\lambda}_{\mu\nu}.
\end{eqnarray}
We take the part describing tensor gravitational waves in perturbations:
\begin{eqnarray}
	\label{tensor perturb}
	& h_{00}=h_{0i}=0,\quad
	 h_{ij}=H_{ij},\quad
	\Sigma^{0}_{00}=\Sigma^{0}_{0i}=\Sigma^{i}_{00}=0,\nonumber
	\\
    &\Sigma^{0}_{ij}=A_{ij},\quad
    \Sigma^{i}_{0j}=B^i_j,\quad
    \Sigma^{i}_{jk}=\partial^{i}C_{jk}+\partial_{(j}D^{i}_{k)}.
\end{eqnarray}
 Here, $H_{ij}, A_{ij}, B_{ij}, C_{ij}$ and $D_{ij}$ are symmetric transverse traceless tensors. They satisfy
\begin{eqnarray}
	\label{symmetric transverse traceless}
	&H_{ij}=H_{ji},~
	A_{ij}=A_{ji},~
	B_{ij}=B_{ji},~
	C_{ij}=C_{ji},~
	D_{ij}=D_{ji},\nonumber
	\\
	&H_{i}^{i}=A_{i}^{i}=B_{i}^{i}=C_{i}^{i}=D_{i}^{i}=0,\\
	&\partial^{i}H_{ij}=\partial^{i}A_{ij}=\partial^{i}B_{ij}=\partial^{i}C_{ij}=\partial^{i}D_{ij}=0.
 \nonumber
\end{eqnarray}
Only in this paragraph, we use $\delta^{ij}$ ($\delta_{ij}$) to raise and lower the index. In Appendix \ref{app: B}, we give the decomposition of the connection and explain why the perturbations describing the tensor gravitational waves are given by Eq.~(\ref{tensor perturb}).

Without losing generality, we consider the propagation direction of gravitational waves as $+z$ direction. At this time, it can be seen from (\ref{tensor perturb}) that the components of the perturbations $h_{\mu\nu}$ and $\Sigma^{\lambda}_{\mu\nu}$ that may not be zero are
\begin{eqnarray}
	\label{h,Sigma,per}
	h_{12},\quad
	h_{11}=-h_{22},\quad
	\Sigma^{0}_{11}=-\Sigma^{0}_{22},\quad
	\Sigma^{0}_{12},\quad
	\Sigma^{1}_{01}=-\Sigma^{2}_{02}, \quad\nonumber
	\\
	\Sigma^{1}_{02}=\Sigma^{2}_{01},\quad
	\Sigma^{1}_{13}=-\Sigma^{2}_{23}, \quad
	\Sigma^{1}_{23}=\Sigma^{2}_{13},\quad
	\Sigma^{3}_{11}=-\Sigma^{3}_{22},\quad
	\Sigma^{3}_{12}.
\end{eqnarray}
By expanding the second-order terms of the perturbations (\ref{h,Sigma,per}) in the action (\ref{action tot}) and varying the action with respect to the perturbations, we can obtain the linear perturbation equations describing the tensor gravitational waves. These equations are very lengthy and easy to obtain, so they are not listed here.

Now we have obtained the linear perturbation equations describing the tensor gravitational waves. Next, we will use the equations to obtain the speed of the tensor gravitational waves.

Before that, we will take metric Horndeski theory as an example to demonstrate how to obtain the speed of tensor gravitational waves from the linear perturbation equation. In metric Horndeski theory, the linear perturbation equation describing the tensor gravitational waves is given by \cite{R.Kase}:
\begin{eqnarray}
	\label{metric Horndeski eq}
	\ddot{h}+b(t)\dot{h}-\frac{{c_t}^2(t)}{a^2(t)}\Delta h=0,
\end{eqnarray}
where $h$ is the component $h_{11}$ or $h_{12}$, $b$ and $c_t$ are functions of time, and $\Delta$ is the Laplace operator. For $h(t,z)$ propagating along the $+z$ direction, we make a Fourier transform:
\begin{eqnarray}
	\label{Fourier transform}
	h(t,z)=\int d^3 k_3 f_{k_3}(t) e^{-ik_{3 z}}.
\end{eqnarray}
By substituting Eq. (\ref{Fourier transform}) into Eq. (\ref{metric Horndeski eq}), and using the linearity of Eq. (\ref{metric Horndeski eq}), we obtain the following equation:
\begin{eqnarray}
	\label{f_k_3}
	\ddot{f}_{k_3}+b(t)\dot{f}_{k_3}+\frac{{c_t}^2(t)}{a^2(t)} k_3^2 f_{k_3}=0.
\end{eqnarray}
This allows us to consider only the case with a single spatial wave vector $k_3$:
\begin{eqnarray}
	\label{h only k3}
	h=f(t) e^{-i {k_3} z},
\end{eqnarray}
where $f(t)$ can always be expressed as
\begin{eqnarray}
	\label{f(t)}
	f(t)=F(t) e^{i {k_0}(t) t}.
\end{eqnarray}
Here, $F$ is the norm of $f(t)$ and ${k_0}(t) t$ is the argument. Therefore, $F$ and ${k_0}$ are real numbers.

Considering that the gravitational wave is observed near time $t_0$ and the observation duration is $\Delta T$, that is, the observation time $t \in [t_0-\frac{\Delta T}{2},t_0+\frac{\Delta T}{2}]$. The duration $\Delta T$ is about the same order of magnitude as the period of the gravitational wave, and during this time, the amplitude and phase of the gravitational wave change very little:
\begin{eqnarray}
	\label{approximate 1}
	\Delta T \sim \frac{2\pi}{k_{0}} \sim \frac{1}{k_{0}}, \quad
	\dot{F}\Delta T \ll F, \quad
	\dot{k_0}\Delta T \ll k_0.
\end{eqnarray}
Thus, $h=F(t)e^{i[{k_0(t) t-k_3 z}]}$ can be approximated as a plane gravitational wave near $t_0$:
\begin{eqnarray}
	\label{plane gravitational wave}
	h=F(t_0)e^{i[k_0(t_0) t-k_3 z]}.
\end{eqnarray}
For the evolution of the cosmic background, the changes of $a,b$ and $c_t$ in Eq. (\ref{metric Horndeski eq}) during this period are also small:
\begin{eqnarray}
	\label{approximate 2}
	\quad \dot{a}\Delta T \ll a, \quad
	\dot{b}\Delta T \ll b, \quad
	\dot{c_t}\Delta T \ll c_t.
\end{eqnarray}
So Eq. (\ref{metric Horndeski eq}) near $t_0$ can be approximated as
\begin{eqnarray}
	\label{approximate metric Horndeski eq}
	\ddot{h}+b(t_0)\dot{h}-\frac{{c_t}^2(t_0)}{a^2(t_0)}\Delta h=0.
\end{eqnarray}
By substituting Eq. (\ref{plane gravitational wave}) into Eq. (\ref{approximate metric Horndeski eq}), we can obtain
\begin{eqnarray}
	\label{k0 k3 demo}
	-{k_0}^2(t_0)+i b(t_0) k_{0}(t_0)+\frac{{c_t}^2(t_0)}{a^2(t_0)} {k_3}^2=0.
\end{eqnarray}
The gravitational waves we can observe have large $k_0$ and $k_3$, which makes the linear term of wave vector component (uniformly recorded as $k$) in the above equation negligible compared with the quadratic term of $k$. Thus, by Eq. (\ref{k0 k3 demo}), the relationship between $k_0$ and $k_3$ will satisfy
\begin{eqnarray}
	\label{k0 k3}
	-{k_0}^2(t_0)+\frac{{c_t}^2(t_0)}{a^2(t_0)} {k_3}^2=0.
\end{eqnarray}
Just write (\ref{plane gravitational wave}) as
\begin{eqnarray}
	\label{plane gravitational wave2}
	h=F(t_0)e^{ik_0(t_0) t}e^{-\left(\frac{k_3}{a(t_0)}\right) \left(a(t_0)z\right)},
	\end{eqnarray}
and using Eq. (\ref{k0 k3}), we can see that the tensor gravitational wave speed $c_g$ at time $t_0$ is
\begin{eqnarray}
	\label{cg}
	c_g(t_0)=\frac{a(t_0)k_0(t_0)}{k_3}=\frac{a(t_0)}{k_3} \frac{c_t(t_0)}{a(t_0)} k_3=c_t(t_0).
\end{eqnarray}
The speed (\ref{cg}) obtained by this method is the same as that of Refs. \cite{R.Kase,T.Kobayashi}.

Similar to the above analysis, for Palatini-Horndeski theory near a certain time, we also approximate the coefficients of perturbations in the linear perturbation equations to constants which are independent of time. In addition, we also approximate the perturbations (\ref{h,Sigma,per}) to the form of plane gravitational waves:
\begin{eqnarray}
	\label{per plane gravitational waves}
	h_{\mu\nu}=\bar{h}_{\mu\nu} e^{i\left(k_0 t-k_3 z\right)},\quad \Sigma^{\lambda}_{\mu\nu}=\bar{\Sigma}^{\lambda}_{\mu\nu} e^{i\left(k_0 t-k_3 z\right)}.
\end{eqnarray}
Here, $\bar{h}_{\mu\nu}$ and $\bar{\Sigma}^{\lambda}_{\mu\nu} $ are amplitudes. Similar to the above example of metric Horndeski theory, by substituting (\ref{per plane gravitational waves}) into the approximated linear perturbation equations, we can obtain the linear equations with amplitudes $\bar{h}_{\mu\nu}$ and $\bar{\Sigma}^{\lambda}_{\mu\nu}$ as the variable. This equations can be written in matrix form:
\begin{eqnarray}
	\label{matrix form}
	AX=0,
\end{eqnarray}
where $A$ is a $10 \times 10$ matrix and it depends on variables $k_0$ and $k_3$. $X$ is a column vector composed of the components of the amplitudes $\bar{h}_{\mu\nu}$ and $\bar{\Sigma}^{\lambda}_{\mu\nu}$. The specific expression of $A$ is very lengthy and easy to obtain, so it is not listed in this paper.

Equation (\ref{matrix form}) has a gravitational wave solution if and only if
\begin{eqnarray}
	\label{det A=0}
	\det (A)=0.
\end{eqnarray}
As in the above example, we consider $k_0$ and $k_3$ to be large. Thus, the lower power terms of $k$ in $\det(A)$ are ignored, and only the highest power terms of $k$ are retained. We mark the remaining quantity in $\det(A)$ as $\mathcal{A}$. Therefore, according to the equation
\begin{eqnarray}
	\label{mathcal{A}}
	\mathcal{A}(k_0,k_3)=0,
\end{eqnarray}
we can know the relationship between $k_0$ and $k_3$, so as to solve the speed of tensor gravitational waves.

Now, we will calculate the speed of tensor gravitational waves propagating in a spatially flat cosmological background. 

We divide the parameter space of Palatini-Horndeski theory into two classes.

\textbf{\emph{Class \uppercase\expandafter{\romannumeral1}}}: $G_{5}(\phi,X)=0$. In this class, by solving Eq. (\ref{mathcal{A}}), we find that the tensor gravitational wave speed $c_g$ is given by
\begin{eqnarray}
	\label{class1 cg}
	c_g^2=\frac{a^2 k_0^2}{k_3^2}=\left(1+\frac{1}{2} \frac{G_{4,X}}{G_4} {\dot{\phi}}^2\right)^2.
\end{eqnarray}
Thus, the condition that the tensor gravitational wave speed is always the speed of light in any spatially flat cosmological background requires
\begin{eqnarray}
	\label{class1 cg condition}
	{G_{4,X}} {\dot{\phi}}^2=0
\end{eqnarray}
for any spatially flat cosmological background.

In this class, by solving the background equation (\ref{BG EQ with P}), we find that $\big(\ddot{a},\dddot{\phi},\Gamma^{0}_{00},{\Gamma}^{1}_{01},{\Gamma}^{0}_{11}\big)$ can be expressed as functions of $\big(a,\dot{a},\phi,\dot{\phi},\ddot{\phi},P,\epsilon\big)$. Further considering the equation of state and the energy conservation equation of the ideal fluid, we can also use $\big(P,a,\dot{a}\big)$ to express $\big(\dot{P},\dot{\epsilon}\big)$. Therefore, as long as we know $\big(a,\dot{a},\phi,\dot{\phi},\ddot{\phi},P,\epsilon\big)$, we can obtain $\big(\ddot{a},\dddot{\phi},\Gamma^{0}_{00},{\Gamma}^{1}_{01},{\Gamma}^{0}_{11},\dot{P},\dot{\epsilon}\big)$. This determines the initial value condition of the background equation (\ref{BG EQ with P}). The specific expressions of these variables are very lengthy and easy to obtain, so we have not listed them.

Considering that there are different equations of state for different types of matters, $P$ and $\epsilon$ can be considered as independent variables. Therefore, the condition that the tensor gravitational wave speed is the speed of light in any spatially flat cosmological background is equivalent to the following condition: at any values of the variables $\big(a,\dot{a},\phi,\dot{\phi},\ddot{\phi},P,\epsilon\big)$, condition (\ref{class1 cg condition}) is always true. This requires that $G_{4,X}$ is always vanishing.

In this way, we find that in Class {\uppercase\expandafter{\romannumeral1}}, only subclass
\begin{eqnarray}
	\label{class1 cg condition fin}
	G_5=0,\quad
	G_{4,X}=0
\end{eqnarray}
satisfies the condition that the tensor gravitational wave speed is the speed of light in any spatially flat cosmological background.

\textbf{\emph{Class \uppercase\expandafter{\romannumeral2}}}: $G_{5}(\phi,X)\neq0$. In this class, by solving the background equation (\ref{BG EQ with P}), we can see that $\big(\dot{a},\dddot{\phi},\dot{\Gamma}^{0}_{00},\dot{\Gamma}^{0}_{11},\dot{\Gamma}^{1}_{01}\big)$ can be expressed as the functions of $\big(a,\phi,\dot{\phi},\ddot{\phi},{\Gamma}^{0}_{00},{\Gamma}^{0}_{11},{\Gamma}^{1}_{01},P,\epsilon\big)$. Further considering the state equation and the energy conservation equation of the ideal fluid, we can also use $\big(P,a,\dot{a}\big)$ to express $\big(\dot{P},\dot{\epsilon}\big)$. Therefore, as long as we know $\big(a,\phi,\dot{\phi},\ddot{\phi},{\Gamma}^{0}_{00},{\Gamma}^{0}_{11},{\Gamma}^{1}_{01},P,\epsilon\big)$, we can solve $\big(\dot{a},\dddot{\phi},\dot{\Gamma}^{0}_{00},
    \dot{\Gamma}^{0}_{11},\dot{\Gamma}^{1}_{01},\dot{P},\dot{\epsilon}\big)$.
This determines the initial value condition of the background equation (\ref{BG EQ with P}). The specific expressions of these variables are very lengthy, so we do not list them. By substituting these expressions into Eq. (\ref{mathcal{A}}) and solving it, we can obtain the tensor gravitational wave speed expressed by the variables $\big(a,\phi,\dot{\phi},\ddot{\phi},{\Gamma}^{0}_{00},{\Gamma}^{0}_{11},{\Gamma}^{1}_{01},P,\epsilon\big)$.

In fact, the tensor gravitational wave speed we solved is not unique in this class, and it has two possible solutions $c_{g1}$ and $c_{g2}$. These two speeds are generally different. However, when the matter pressure $P=0$, we have $c_{g1}=c_{g2}$. The specific expressions of $c_{g1}$ and $c_{g2}$ are very lengthy, so we do not list them.

The first speed $c_{g1}^2$ can be expressed as a fraction
 \begin{eqnarray}
 	\label{class2 cg1}
 	c_{g1}^2=\frac{\mathcal{N}}{\mathcal{D}}.
 \end{eqnarray}
 It can be seen that $c_{g1}=1$ is equivalent to the numerator part $\mathcal{N}$ on the right side of Eq. (\ref{class2 cg1}) minus the denominator part $\mathcal{D}$ equal to $0$:
 \begin{eqnarray}
 	\label{cg1 numerator-denominator}
 	M\equiv{\mathcal{N}}-{\mathcal{D}}=0.
 \end{eqnarray}
By expanding the brackets, $M$ can be expressed as a polynomial about the variables $\big(a,\phi,\dot{\phi},\ddot{\phi},{\Gamma}^{0}_{00},{\Gamma}^{0}_{11},{\Gamma}^{1}_{01},P,\epsilon\big)$. If we require the tensor gravitational wave speed $c_{g1}$ to be the speed of light under any spatially flat  cosmological background, then for any values of the variables $\big(a,\phi,\dot{\phi},\ddot{\phi},{\Gamma}^{0}_{00},{\Gamma}^{0}_{11},{\Gamma}^{1}_{01},P,\epsilon\big)$, this polynomial should be $0$. We notice that in this polynomial, the term where
$(\Gamma^{0}_{00})^4\epsilon$ appears is
\begin{eqnarray}
	\label{cg1 000^4}
    6 a^{14} (G_5)^5  {\dot{\phi}}^5
    \Big( 5  G_{5,X} {\dot{\phi}}^2  -2  {G_5} \Big) (\Gamma^{0}_{00})^4\epsilon.
\end{eqnarray}
Therefore, the above condition requires
\begin{eqnarray}
	G_5=0 ~~\text{or}~~G_5=\frac{5}{2}{\dot{\phi}}^2G_{5,X}. \label{CassIIcondition1}
\end{eqnarray}
If substituting the condition $G_5=\frac{5}{2}{\dot{\phi}}^2G_{5,X}$ into Eq. (\ref{cg1 numerator-denominator}), we again notice that in this polynomial, the term where $({\Gamma^{1}_{01}})^4\epsilon$ appears is
\begin{eqnarray}
	\label{cg1 101^4}
	-\frac{1171875}{16} a^{14} (G_{5,X})^6 {\dot{\phi}}^{17} (\Gamma^{1}_{01})^4\epsilon.
\end{eqnarray}
Therefore, the above condition further requires $G_{5,X}=0$. Combining with the condition (\ref{CassIIcondition1}) we have $G_5=0$. However, this is inconsistent with the assumption $G_5 \neq 0$ in this class.

For the second solution $c_{g2}$, using the same analysis method as that used to analyze the first solution $c_{g1}$, we find that the condition of $c_{g2}=1$ also requires $G_5=0$.

To sum up, for Palatini-Horndeski theory, the parameter space satisfying the condition that the tensor gravitational wave speed is the speed of light under any spatially flat cosmological background is only $G_{4,X}=0$ and $G_5=0$.

\section{Conclusion}
\label{sec: 5}
In this paper, 
we calculated the speed of tensor gravitational waves 
in the spatially flat cosmological background. It is worth noting that we found that there are two possible speeds of tensor gravitational waves in Class \uppercase\expandafter{\romannumeral2}. This is due to the additional degrees of freedom introduced by the tensor perturbations of the connection. It seems to imply that if we observe two tensor gravitational waves with different speeds in the future, the theory of gravitation describing our world may be described by the Palatini formalism.

However, if we further require the tensor gravitational wave speed to be the speed of light $c$ in any spatially flat cosmological background, then only
 \begin{eqnarray}
 \label{action fin}
 S\left(g,\Gamma,\phi\right) = \int d^4x \sqrt{-g}\big[K(\phi,X)-G_{3}(\phi,X){\tilde{\Box}}\phi+G_{4}(\phi)\tilde{R} \big]
 \end{eqnarray}
 is left as the possible action in the above two subclasses of Palatini-Horndeski theory. Reference \cite{Helpin} pointed out that the action (\ref{action fin}) in the Palatini formalism is actually equivalent to the following action in the metric formalism:
 \begin{eqnarray}
 	\label{KGB}
 	S\left(g,\phi\right) = \int d^4x \sqrt{-g}\big[\bar{K}(\phi,X)-G_{3}(\phi,X){{\Box}}\phi+G_{4}(\phi){R}\big].
 \end{eqnarray}
 Here,
 \begin{eqnarray}
 	\bar{K}=K+\left(-2G_3 G_{4,\phi}+3 G_{4,\phi}^2-\frac{2}{3} G_3^2\right)\frac{X}{G_4}.
 \end{eqnarray}
It can be seen that the action (\ref{action fin}) in the Palatini formalism actually still belongs to metric Horndeski theory. Therefore, the parameter space of Palatini-Horndeski theory compatible with GW170817 does not have the Ostrogradsky instability. It should be noted that the action (\ref{KGB}) is the only subclass of metric Horndeski theory that is compatible with the condition that the tensor gravitational wave speed is the speed of light $c$ in any spatially flat cosmological background.

Finally, it should be pointed out that this does not mean that the scalar-tensor gravity in the Palatini formalism must not beyond the framework of metric Horndeski theory. It is because Palatini-Horndeski theory considered in this paper is not the most general theory of scalar-tensor gravity in the Palatini formalism. A more general discussion needs to study more general action, which needs to be studied in future work.

\section*{Acknowledgments}
We would like to thank Yu-Peng Zhang for useful discussion. This work is supported in part by the National Key Research and Development Program of China (Grant No. 2020YFC2201503), the National Natural Science Foundation of China (Grants No. 11875151 and  No. 12047501), the 111 Project (Grant No. B20063), the Department of education of Gansu Province: Outstanding Graduate ``Innovation Star" Project (Grant No. 2022CXZX-059) and Lanzhou City's scientific research funding subsidy to Lanzhou University.

\appendix

\section{Decomposition of connection}
\label{app: B}
In this appendix, we use $\delta^{ij}$ ($\delta_{ij}$) to raise and lower the index. Therefore, it will not cause ambiguity if the upper and lower indices of a tensor are not distinguished.

We can decompose the perturbation of the connection $\Sigma^{\lambda}_{\mu\nu}$ into the following forms:
\begin{eqnarray}
\Sigma^{0}_{00}&=&\Sigma^{0}_{00}.
\end{eqnarray}
\begin{eqnarray}
\Sigma^{i}_{00}&=&\partial_{i}F+G_i,\quad  \Sigma^{0}_{0i}=\partial_{i}M+N_i.
\end{eqnarray}
\begin{eqnarray}
\Sigma^{0}_{ij}&=&A\delta_{ij}+\partial_{i}\partial_{j}B+\partial_{i}C_j+\partial_{j}C_i+S_{ij},\nonumber
\\
\Sigma^{i}_{0j}&=&\bar{A}\delta_{ij}+\partial_{i}\partial_{j}\bar{B}+\partial_{i}\bar{C}_j+\partial_{j}\bar{D}_i+U_{ij}+V_{ij}.
\end{eqnarray}
\begin{eqnarray}
	\Sigma^{i}_{jk}&=&B^{i}_{jk}+\partial_{i}C_{jk}+\partial_{(j}D_{k)i}+\partial_{(j}E_{k)i}+\partial_{i}\partial_{(j}f_{k)}+\partial_{j}\partial_{k}g_i\nonumber
	\\
	&+&\partial_{i}\partial_{j}\partial_{k}l+h_i\delta_{jk}+q_{(j}\delta_{k)i}+\left(\partial_{i}m\right)\delta_{jk}+\left(\partial_{(j}n\right)\delta_{k)i}.
\end{eqnarray}
Here,
\begin{eqnarray}
	\partial_{i}G^i=\partial_{i}N^i=\partial_{i}C^i=\partial_{i}{\bar{C}}^i=\partial_{i}{\bar{D}}^i=\partial_{i}f^i=\partial_{i}g^i=\partial_{i}h^i=\partial_{i}q^i=0.
\end{eqnarray}
\begin{eqnarray}
	S_{i}^{i}=U_{i}^{i}=V_{i}^{i}=C_{i}^{i}=D_{i}^{i}=E_{i}^{i}=0;\nonumber
		\\
     \partial_{i}S^{i}_{j}=\partial_{i}U^{i}_{j}=\partial_{i}V^{i}_{j}=\partial_{i}C^{i}_{j}=\partial_{i}D^{i}_{j}=\partial_{i}E^{i}_{j}=0;\nonumber
		\\
	S_{ij}=S_{ji},\quad U_{ij}=U_{ji},\quad C_{ij}=C_{ji},\nonumber
	\\
	 D_{ij}=D_{ji},\quad V_{ij}=-V_{ji},\quad E_{ij}=-E_{ji}.
\end{eqnarray}
\begin{eqnarray}	
B^{i}_{jk}=	B^{i}_{kj},\quad B^{i}_{ik}=B^{i}_{jk}\delta^{jk}=0,\quad \partial_{i}B^{i}_{jk}=\partial^{j}B^{i}_{jk}=0.
\end{eqnarray}
The notations used in the appendix are not related to those in the text and should not be confused. The method to prove that the perturbation of the connection $\Sigma^{\lambda}_{\mu\nu}$ can always be decomposed into the above form is the same as the method for the perturbation of the metric $h_{\mu\nu}$.

Similarly, we can decompose the linear perturbation equations into several sets of coupled equations. The perturbation describing the tensor gravitational waves is the transverse traceless tensor part of the metric perturbation $h^{TT}_{\mu\nu}$. Consider that each term in the linear perturbation equations is a combination of $\delta_{ij}$, $\partial_{i}$, a time dependent function and a perturbation, and since $h^{TT}_{\mu\nu}$ is a transverse traceless symmetric tensor, only $S_{ij}$, $U_{ij}$, $C_{ij}$ and $D_{ij}$ are coupled with $h^{TT}_{\mu\nu}$ in the perturbations of the connection (A1-A4). It is the reason why we take the perturbations (\ref{tensor perturb}) in the text.

\end{document}